\documentclass[]{aa} 

\usepackage{graphicx}
\usepackage{color}
\usepackage{float}

\begin{document}

\title{Magnetic fields during galaxy mergers}
\titlerunning{Magnetic fields during galaxy mergers} 
\authorrunning{Rodenbeck \& Schleicher}

\author
  {Kai Rodenbeck\inst{1}
\and
Dominik R.\,G. Schleicher
  \inst{2}
  }

\institute{ Max-Planck-Institute for Solar System Research, Justus-von-Liebig-Weg 3, 37077 G\"ottingen, Germany; \newline \email{rodenbeck@mps.mpg.de}
\and
Departamento de Astronom\'ia, Facultad Ciencias F\'isicas y Matem\'aticas, Universidad de Concepci\'on, Av. Esteban Iturra s/n Barrio Universitario, Casilla 160-C, Concepci\'on, Chile; \email{dschleicher@astro-udec.cl}
}

\date{\today}


\abstract{
Galaxy mergers are expected to play a central role for the evolution of galaxies, and may have a strong impact on their magnetic fields. We present the first grid-based 3D magneto-hydrodynamical simulations investigating the evolution of magnetic fields during merger events. For this purpose, we employ a simplified model considering the merger event of magnetized gaseous disks in the absence of stellar feedback {and without a stellar or dark matter component}. We show that our model naturally leads to the production of two peaks in the evolution of the average magnetic field strength within $5$~kpc, within $25$~kpc and on scales {in between} $5$ and $25$~kpc. The latter is consistent with the peak in the magnetic field strength reported by \citet{Drzazga11} in  a merger sequence of observed galaxies. We show that the peak on the galactic scale and in the outer regions is likely due to geometrical effects, as the core of one galaxy enters the outskirts of the other one. In addition, there is a physical enhancement of the magnetic field within the central $\sim5$~kpc, reflecting the enhancement in density due to efficient angular momentum transport. We conclude that high-resolution observations of the central regions will be particularly relevant to probe the evolution of magnetic field structures during merger events.
}

\maketitle

\section{Introduction}

Galaxy mergers are central for the evolution of galaxies, and have a strong influence on many of their properties. In a pioneering study to investigate the impact of mergers, \citet{Toomre77} has proposed a merging sequence of 11 peculiar galaxies spanning a range of pre- and post-mergers that became known as the {\em Toomre sequence}. This sequence suggested that the final product of mergers is likely to resemble an elliptical galaxy. This behavior was confirmed by  numerical simulations following the N-body dynamics of the stellar population, finding the characteristic formation of $r^{1/4}$ profiles observed in elliptical galaxies \citep{Barnes88, Barnes92, Hernquist92, Hernquist93, Privon13}. Observational studies have provided evidence for this scenario, reporting a characteristic $r^{1/4}$ profile in merger remnants \citep{Joseph85} and finding low surface brightness loops and shells in elliptical galaxies,  which are likely the remnants from previous mergers \citep{Malin80, Schweizer88}. \citet{Forbes98} have shown that late-stage disk-disk mergers and ellipticals with young stellar populations deviate from the fundamental plane of elliptical galaxies due to a centrally located starburst induced by the merger event.
  
Gravitational instabilities acting during the merger event may channel significant gas flows towards the center of the galaxy, enhancing the activity of star formation or of a central supermassive black hole \citep{Mihos96}. High molecular gas fractions have  been observed in IRAS starburst galaxies, which are considered as gas-rich systems close to the final stages of merging \citep{Sanders96, Planesas97, Kennicutt98}. The enhanced star-formation activity has been confirmed both in the far-infrared \citep{Kennicutt98} as well as in the radio \citep{Hummel81, Hummel90} and the optical \citep{Keel85}. Near-infrared observations with the Hubble Space Telescope show a distinct trend for the nuclei to become more luminous in more advanced merging stages \citep{Rossa07}. Similarly, the fraction of interacting systems in IRAS-selected samples increases with far-infrared luminosity \citep{Lawrence89, Gallimore93}.

 \citet{Gao99} have used the projected separation between the nuclei of merging galaxies as an estimator for the interaction stage, finding clear evidence for increasing star formation efficiencies, defined by the ratio of far-infrared luminosity to molecular hydrogen mass, with decreasing spatial separation. The latter potentially indicates more efficient star formation in higher-density gas. The increase of star formation activity during the merger event is also reflected in X-ray observations \citep{Read98, Brassington07} and in the UV \citep{Smith10}. \citet{Hibbard96} investigated the cold gas properties from a small sample of pre- and post-mergers, finding strong differences in the HI distribution in pre- and post-mergers, with increasing fractions of HI outside the optical bodies towards the later stages. They propose that the latter is due to the ejection of cold gas in tidal features. 
  
  \citet{Georgakakis00} investigated a sample of $53$ interacting galaxy pairs, combining radio observations with optical data by \citet{Keel95}, far-infrared data by \citet{Gao99} and $60$~$\mu$m data from the sample of \citet{Surace93} based on the IRAS Bright Galaxy Sample \citep{Soifer87}. They defined the interaction parameter in terms of the effective age before or after the merger, confirming a significant increase of the star formation efficiency during the merger event. The latter is reflected in a correlation between the star formation efficiency and the H$_2$ surface density. The amount of molecular gas is found to increase during the merger as well, and to fall off after the merger. Beyond the statistical samples, some interacting galaxy pairs have been studied in particular detail. For instance, the Antennae galaxies (NGC~4038/39) have been explored with Chandra in the X-rays \citep{Fabbiano01}, with Herschel-PACS revealing obscured star formation \citep{Klaas10} and with the Submillimeter Array pursuing high-resolution CO observations \citep{Ueda12}. The change of the gas distribution due to the merger event may also be relevant for the cosmological growth of black holes \citep{Mayer10, Treister12, Ferrara13, Capelo15}.

In addition to star formation and the molecular gas, magnetic fields are a relevant component of galaxies, which can influence the star formation rate \citep[e.g.][]{Banerjee09, Vazquez11, Federrath12}. The structure of the galactic magnetic field has been studied in detail in nearby galaxies \citep[e.g.][]{Beck99, Beck05, Fletcher11, Tabatabaei13, Heesen14}. More recently, the presence of magnetic fields was confirmed also in nearby dwarf galaxies \citep{Chyzy00, Kepley10, Heesen11, Kepley11, Chyzy11}. The far-infrared - radio correlation suggests that the presence of magnetic fields is strongly correlated with star formation activity, where the magnetic field strength $B$ scales approximately as the star formation surface density $\Sigma_{\rm SFR}$ to the one third \citep{deJong85, Helou85, Niklas97, Yun01, Basu12, Schleicher13c}, and has also been confirmed at high redshift \citep{Murphy09, Jarvis10, Ivison10, Bourne11, Casey12}.

  While radio data have been available for interacting galaxy pairs early on \citep[e.g.][]{Condon83, Condon93, Condon02}, the first detailed analysis of the magnetic field structure in an interacting galaxy pair has been pursued by \citet{Chyzy04} for the Antennae galaxies. With $\sim20$~$\mu$G, the mean total magnetic field strength is significantly greater than in isolated spirals, and could be enhanced via interaction-induced star formation. The field regularity appears to be lower compared to isolated galaxies, reflecting additional distortions or feedback related to star formation. The observed structures are roughly consistent with magneto-hydrodynamical (MHD) simulations by \citet{Kotarba10} based on a smoothed particle hydrodynamics (SPH) aproach. The galaxy pair NGC~2207/IC~2163 has been observed in the optical by \citet{Elmegreen95a}, and was modeled with N-body simulations by \citet{Elmegreen95b} reproducing the main dynamical features. The radio emission at $4.86$~GHz covers the optical extent of the system, but includes a hole in the central part of NGC~2207 \citep{Condon83}.  For the Taffy (UGC~12914/15), a galaxy pair with a nearly head-on encounter, a significant amount of data is available, including radio and HI \citep{Condon93} as well as CO and the dust continuum \citep{Gao03, Zhu07}. A similar pair of post-collision spiral galaxies, UGC~813/16 was detected at $1.40$~GHz and forms the so-called Taffy2 \citep{Condon02}. 

  The first more systematic study of magnetic fields in a complete merging sequence has been pursued by \citet{Drzazga11}. For this purpose, they compiled a sample of data from the Very Large Array (VLA) based on the merging sequence of \citet{Toomre77} and \citet{Brassington07} as well as other interacting system, leading to a total of $24$ interacting galaxy pairs, including pre- and post-merger stages. While they do not consider them as a complete sample of colliding galaxies, they do represent the different stages of tidal interaction and merging.  For this sample, a correlation of the average magnetic field strength in the whole galaxy and in the central region is found with the star formation surface density $\Sigma_{\rm SFR}$. The central magnetic field strength scales as $\Sigma_{\rm SFR}^{0.27\pm0.03}$, consistent with the scaling relations in spiral galaxies \citep{Niklas97}, and in fact \citet{Drzazga11} report that their sample lies on the far-infrared - radio correlation for spiral galaxies.

The evolution of magnetic fields during merger events is also interesting from a theoretical point of view, as the merger process leads to the production of turbulence, which can enhance both the small-scale magnetic fields via the small-scale dynamo \citep[e.g.][]{Scheko02, Cho09, FederrathPRL, Schober12a, Schleicher13b,Bhat14, Schober15}, and also affect the formation of a large-scale field via dynamo processes \citep{Ruediger93, Brandenburg05, Gressel13, Park13}. The number of numerical simulations exploring such merger scenarios is however still limited. Beyond the MHD-SPH simulations of the Antennae galaxies pursued by \citet{Kotarba10}, the impact of multiple mergers has been explored with the same approach by \citet{Kotarba11}. Magnetic field amplification during minor galaxy mergers has been investigated with a similar approach by \citet{Geng12}. Grid-based simulations of merging systems have been pursued in a 2D-approximation by \citet{Moss14}. However, as turbulence is essentially a 3D phenomenon, it is desireable to explore the evolution of magnetic fields during merger events with three-dimensional grid-based techniques. 

In this paper, we present a toy model to explore 3D effects during the merger of magnetized gaseous disks. {While this model still neglects the impact of the dark matter on the merger events, it shows that geometric projection events can be rather important when interpreting large-scale averages in observations of magnetic fields. We indeed show via a parameter study that the occurence of characteristic peaks due to projection effects is rather insensitive to the details of the orbit.} Our simulation setup is presented in section~\ref{setup} and the main results are given in section~\ref{results}. We show that our simulations can already reproduce some of the main features in the merger sequence by \citet{Drzazga11}. Our results and their implications for the interpretation of observational data are discussed in section~\ref{discussion}.

\section{Simulation setup}\label{setup}
Our simulations are pursued with the publicly available magneto-hydrodynamics (MHD) adaptive-mesh-refinement (AMR) code Enzo\footnote{Webpage Enzo: http://enzo-project.org/} \citep{Shea04, Bryan14}. To solve the MHD equations, we employ the Monotonic Upstream-Centered Scheme for Conservation Laws (MUSCLE) \citep{Leer79} with the divergence-cleaning scheme developed by \citet{Dedner02}. The latter was implemented in the Enzo code by \citet{Wang09}. In this approach, the divergence of $\vec{B}$ is kept small via an additional damping term. The latter is slightly less accurate than the constrained transport approach employed by \citet{Collins10}, but computationally more efficient, which is highly beneficial for the merger simulations pursued here.

In our numerical setup, we consider the collision of two magnetized gaseous disks as a toy model to investigate the evolution of magnetic fields during merger events. In particular, we neglect here the potential impact of a dark matter halo and the stellar component. While the latter provides an external potential which can potentially stabilize the disk, we expect that the qualitative evolution of the magnetic fields will not be strongly depend on their effects. {Similarly, the additional presence of a dark matter halo would stabilize the galaxy against tidal effects and increase both the gravitational and the inertial mass, while the dynamical friction due to the dark matter is likely reduced compared to the baryons.} The initial conditions of the disks are based on the isolated disk models by \citet{Wang10} and \citet{Braun14}. 

\subsection{Gas distribution}

We consider an exponential gaseous disk where the surface density is given as\begin{equation}
\Sigma(r)=\Sigma_0 \mathrm{exp}\left(-\frac{r}{R}\right),\label{sigma}
\end{equation}
with $R$ the characteristic length scale of the disk and $\Sigma_0$ the surface density in the center. For the vertical density profile, we consider a distribution\begin{equation}
\rho(r,z)=\rho(r,z=0)\mathrm{sech}^2\left(\frac{r}{H(r)}  \right),\label{dens}
\end{equation}
where $H(r)$ is the scale height of the disk at radius $r$. In the following, we adopt the abbreviation $\rho_0(r)=\rho(r,z=0)$. Assuming that the self-gravity of the disk is balanced by gas pressure, the scale height is given as\begin{equation}
H(r)=\sqrt{\frac{(1+\epsilon_{\rm mag})c_s^2(r)}{2\pi G\rho_0(r)}},\label{height}
\end{equation}
where $G$ is the gravitational constant, $c_s(r)$ the local sound speed and $\epsilon_{\rm mag}$ denotes the ratio of magnetic to thermal pressure, which is assumed to be constant in the initial condition. We will in the following assume a constant gas temperature in the initial condition and thus also a constant sound speed $c_s$. Through an integration of Eq.~\ref{sigma}, the total mass in the disk is given as
\begin{equation}
M_{\rm disk}=\int_0^\infty 2\pi\Sigma_0 r\,\mathrm{exp}\left(- \frac{r}{R} \right)dr=2\pi\Sigma_0 R^2.
\end{equation}
For a given mass of the disk, the normalization constant $\Sigma_0$ thus follows as\begin{equation}
\Sigma_0=\frac{M_{\rm disk}}{2\pi R^2}.\label{sigma0}
\end{equation}
For the disk density profile in Eq.~(\ref{dens}), we have the following identity:
\begin{equation}
\Sigma(r)=2H(r)\rho_0(r).\label{id}
\end{equation}
Comparing Eqs.~(\ref{sigma}) and (\ref{id}) and inserting the expression (\ref{sigma0}), one can show that\begin{equation}
H(r)=\frac{(1+\epsilon_{\rm mag})c_s^2(r)R^2}{GM_{\rm disk}}\mathrm{exp}\left( \frac{r}{R} \right).
\end{equation}
As a result, the density distribution is given as\begin{equation}
\rho(r,z)=\frac{G}{8\pi (1+\epsilon_{\rm mag})c_s^2}\left( \frac{M_{\rm disk}}{R^2} \right)^2\mathrm{exp}\left(-\frac{2r}{R}  \right)\mathrm{sech}^2\left( \frac{z}{H(r)} \right).
\end{equation}
While the above considerations assume a balance between gravity and thermal pressure, the approach given by \citet{Wang10} can be extended for additional non-thermal pressure components, including magnetic pressure. For this purpose, we adopt the following system of equations:\begin{eqnarray}
\frac{d^2\Phi}{dz^2}&=&4\pi G\rho(r,z),\\
\rho(r,z)&=&\rho_0(r)\mathrm{exp}\left( \Phi_z(r,z)\left[ \left( \frac{d\rho}{dz} \right)^{-1}\frac{dp}{dz} \right]^{-1} \right),\\
\rho_0(r)&=&\frac{\Sigma_0 \mathrm{exp}\left( -\frac{r}{R} \right)}{2\int_0^\infty \mathrm{exp}\left( \Phi_z(r,z)\left[ \left( \frac{d\rho}{dz} \right)^{-1} \frac{dp}{dz} \right]^{-1} \right)dz}.
\end{eqnarray}
with $\Phi_z=\Phi(r,z)-\Phi(r,0)$. This system of equations is solved via an iterative approach, starting with an initial guess for $\rho_0(r)$ based on the solution neglecting the magnetic pressure. The method typically converges within a few iterations.

\subsection{Velocity field}
For galactic disks in which the surface density $\Sigma(r)$ follows an exponential profile, there is an analytical solution describing the case where the gravitational force is completely balanced by rotation \citep{Freeman70}:\begin{equation}
V^2_{\rm grav}(r)=\frac{\pi G\Sigma_0 r^2}{R}\left(I_0\left( \frac{r}{2R} \right) K_0\left( \frac{r}{2R} \right)-I_1\left( \frac{r}{2R} \right) K_1\left( \frac{r}{2R} \right) \right),
\end{equation}
where $I_i$ and $K_i$ are the modified Bessel functions of order $i$. In a realistic disk, the gravitational force is however not only balanced by the centrifugal force, but also by thermal and magnetic pressure. To take this into account, the rotational velocity is corrected for the contribution of thermal and magnetic pressure via \citep{Wang10}\begin{eqnarray}
V^2_{\rm rot}(r)&=&V_{\rm grav}^2(r)-V_{\rm pr}^2(r),\\
V_{\rm Pr}^2(r)&=&2(1+\epsilon_{\rm mag})c_s^2\frac{r}{R}.
\end{eqnarray}

\subsection{Magnetic field}
For simplicity, we assume here that the magnetic field is initially toroidal, and that the magnetic pressure $B^2/(8\pi)$ corresponds to a fixed fraction $\epsilon_{\rm mag}$ of the thermal pressure $\rho c_s^2$. We then have\begin{equation}
B_{\rm tor}=\sqrt{8\pi\epsilon_{\rm mag}\rho c_s^2}.
\end{equation}
In the initial condition, the ratio of the magnetic to thermal pressure will in particular influence the scale height of the disk, the gas distribution and the rotational velocity. 

\begin{figure*}[htbp]
\begin{center}
\includegraphics[scale=0.43]{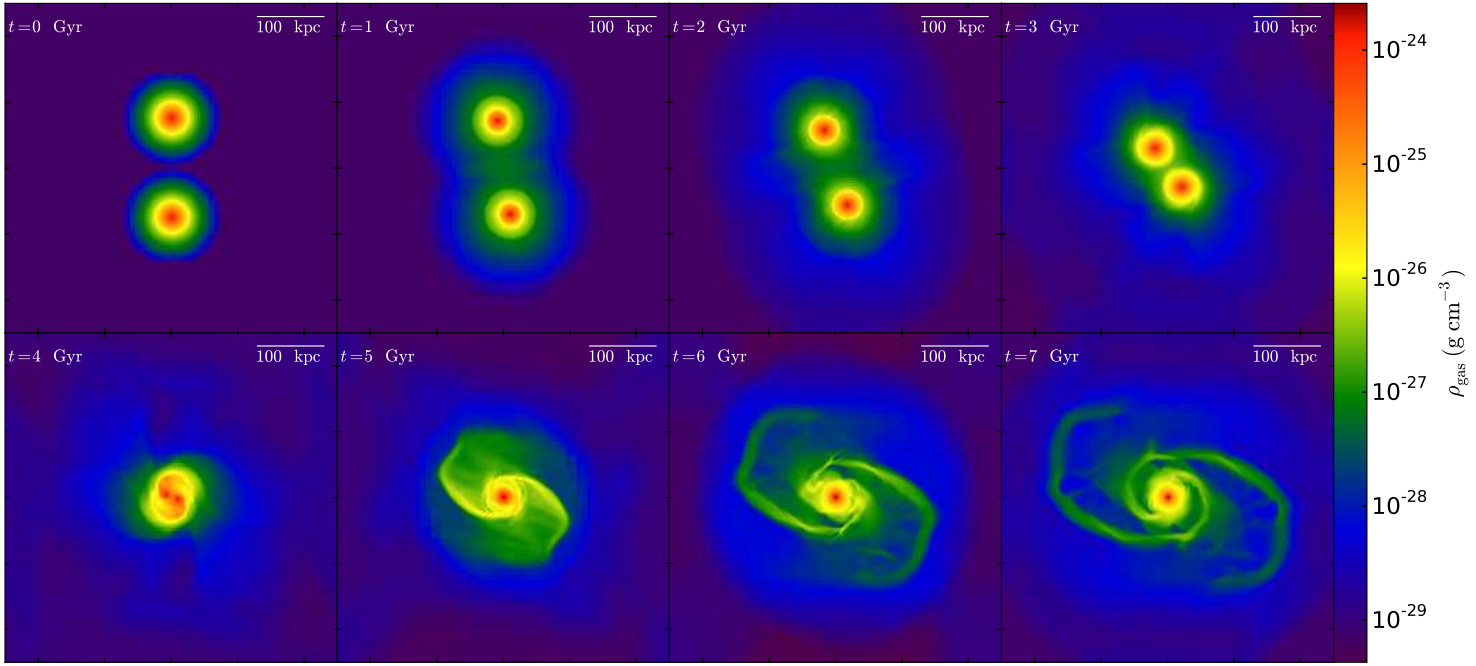}
\caption{Density slice in the x-y plane at different times showing the global evolution in the simulation box {(run A)}, including the merger stage at about $4$~Gyrs and the post-merger configuration after $5$~Gyrs.}
\label{global}
\end{center}
\end{figure*}

\begin{figure*}[htbp]
\begin{center}
\includegraphics[scale=0.43]{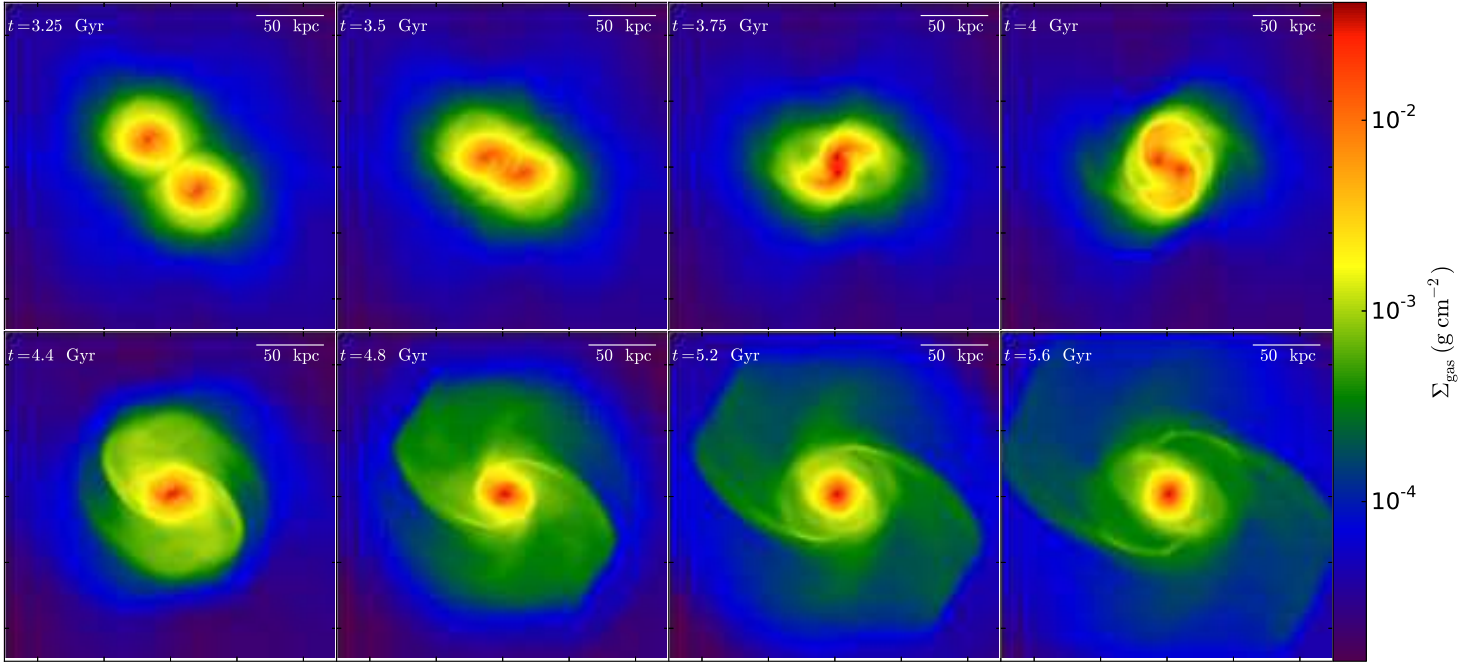}
\caption{Gas density projections between $3.25$~Gyrs and $5.6$~Gyrs, showing the evolution on scales of $250$~kpc around the merger event {(run A)}. The galaxy cores are clearly merged at a time of $4.4$~Gyrs.}
\label{sigma}
\end{center}
\end{figure*}

\begin{figure*}[htbp]
\begin{center}
\includegraphics[scale=0.43]{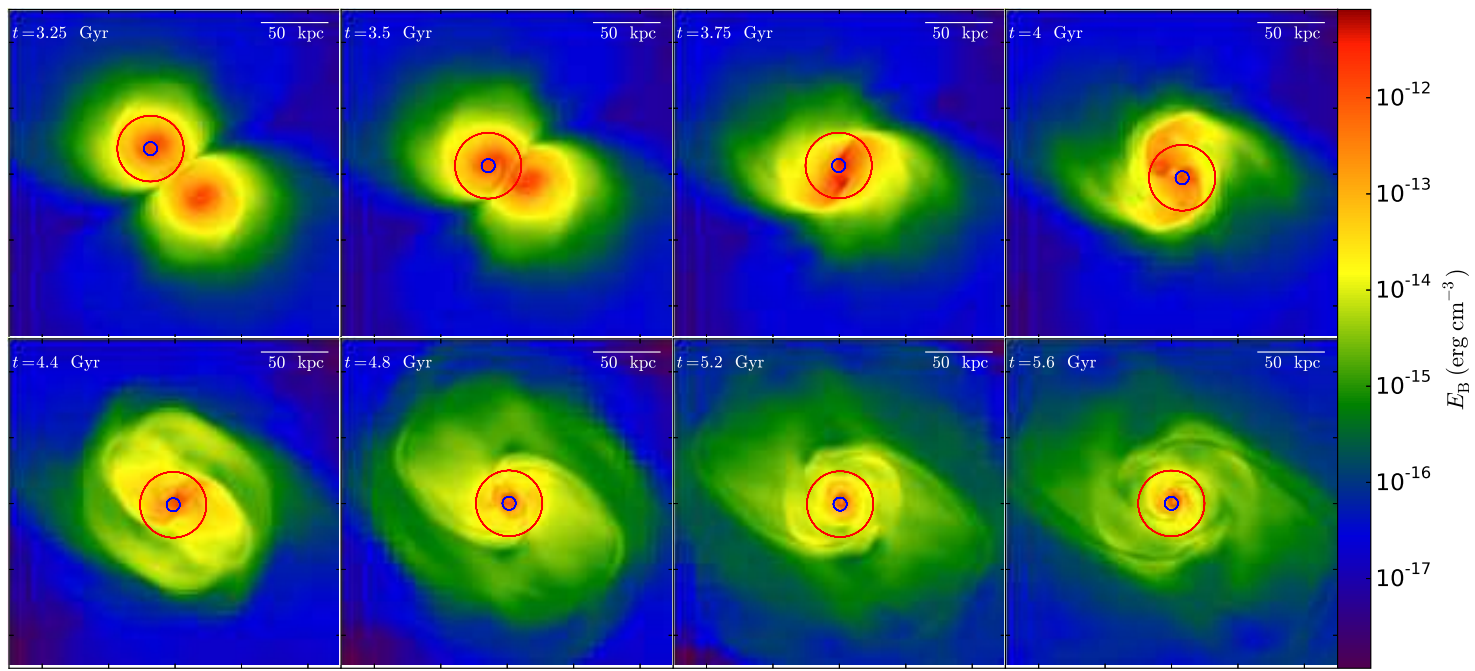}
\caption{Density-weighted projections of the magnetic energy density between $3.25$~Gyrs and $5.6$~Gyrs, showing the evolution on scales of $250$~kpc around the merger event {(run A)}. The structure visible in the magnetic energy resembles the structure visible in the gas density. The galaxy cores have clearly merged after $4.4$~Gyrs. With circles, we indicate the region over which the field strength of the whole galaxy is averaged ($25$~kpc), as well as the region corresponding to an average over the center ($5$~kpc). The circles are centered on the density peak in the simulation.}
\label{Bdens}
\end{center}
\end{figure*}

\subsection{Simulation box}
Overall, we consider a cubic simulation box of size $500$~kpc with periodic boundary conditions, containing two {idealized disks} of equal mass $0.5\times10^{11}$~M$_\odot$ which are aligned with the x-y plane. They have a scale radius of $10$~kpc and a  separation of $150$~kpc, aligned with the y-axis. For the intergalactic medium, we adopt a mean density of $\sim7\times10^{-30}$~g~cm$^{-3}$. The gas {in the disks and in the intergalactic medium} is assumed to be isothermal at a temperature of $5\times10^4$~K. We adopt a ratio of magnetic to thermal pressure of $\epsilon_{\rm mag}=0.5$. {The magnetic field in the intergalactic medium is randomly orientated, with an rms value of $10^{-25}$~G. The medium outside the galaxies is thus strongly dominated by the thermal pressure. Potential divergences of the magnetic field are thus of low magnitude and taken care for via divergence cleaning.}

The  relative velocity of the galaxies is initially aligned with the x-direction and corresponds to $20$~km/s (each galaxy moving with half of that velocity in opposite directions). The velocity is chosen here for practical purposes, as larger initial velocities require a considerably larger simulation box. 

To resolve both the initial structures as well as their subsequent evolution, we employ the adaptive mesh refinement technique, with a top grid resolution of $32^3$ and 6 additional refinement levels, each increasing the resolution by a factor of $2$. The latter corresponds to an effective resolution of $0.24$~pc in the simulation. A new refinement level is introduced both in the initial condition as well as during the dynamical evolution whenever the Jeans length is resolved by less than $32$ cells, so that the internal disk structure is well-resolved. The simulation is evolved for $10$~Gyrs to follow the dynamical evolution.

\begin{figure*}[htbp]
\begin{center}
\includegraphics[scale=0.7]{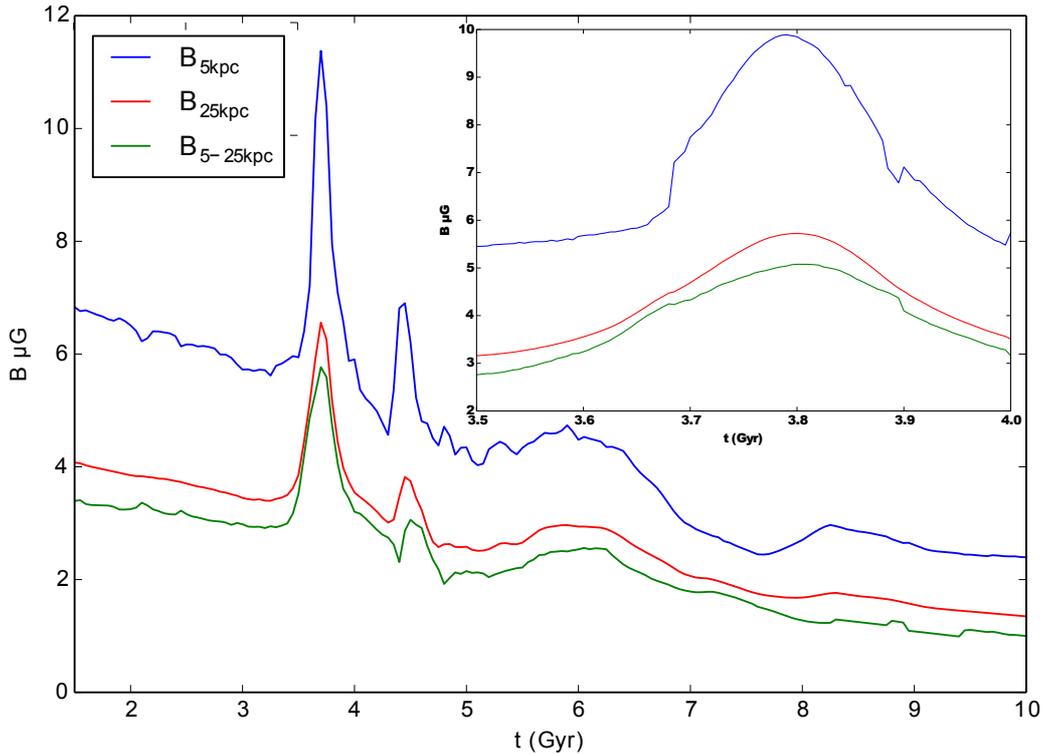}
\caption{The evolution of the average magnetic field strength within the central $5$~kpc, within the central $25$~kpc and on the scales between $5$ and $25$~kpc as a function of time {(run A)}.}
\label{av}
\end{center}
\end{figure*}

\section{Results}\label{results}
{In the following, we will first describe our reference run in greater detail, to show the main qualitative features that may occur. We will in particular see from the reference run that geometric effects are sufficient to explain the qualitative behavior that has been observed. Subsequently, we will show additional runs with varying parameters to address possible variations that may occur.} {A list of all simulations presented here is given in Table~\ref{sims}, the reference run corresponds to run A in the list.}

\begin{table}[htp]
\begin{center}
\begin{tabular}{c|c|c|c|c}
Simulation & Mass ratio & rel. orientation & $v$ & $N_B$\\ \hline
A & 1:1 & $0^\circ$ & $20$ & $2$\\
B & 1:1 & $90^\circ$ & $20$ & $2$\\
C & 1:3 & $0^\circ$ & $20$ & $1$\\
D & 1:1 & $0^\circ$ & $10$ & $1$\\
E & 1:1 & $0^\circ$ & $15$ & $2^*$
\end{tabular}
\end{center}
\label{sims}
\caption{{List of simulations presented here. $v$ denotes the relative velocity in km/s between the idealized disks, and $N_B$ the number of characteristic peaks in the time evolution of the magnetic field on scales of $25$~kpc. The symbol $*$ denotes that the second peak is strongly reduced in its magnitude.}}
\end{table}%

\begin{figure*}[htbp]
\begin{center}
\includegraphics[scale=0.65]{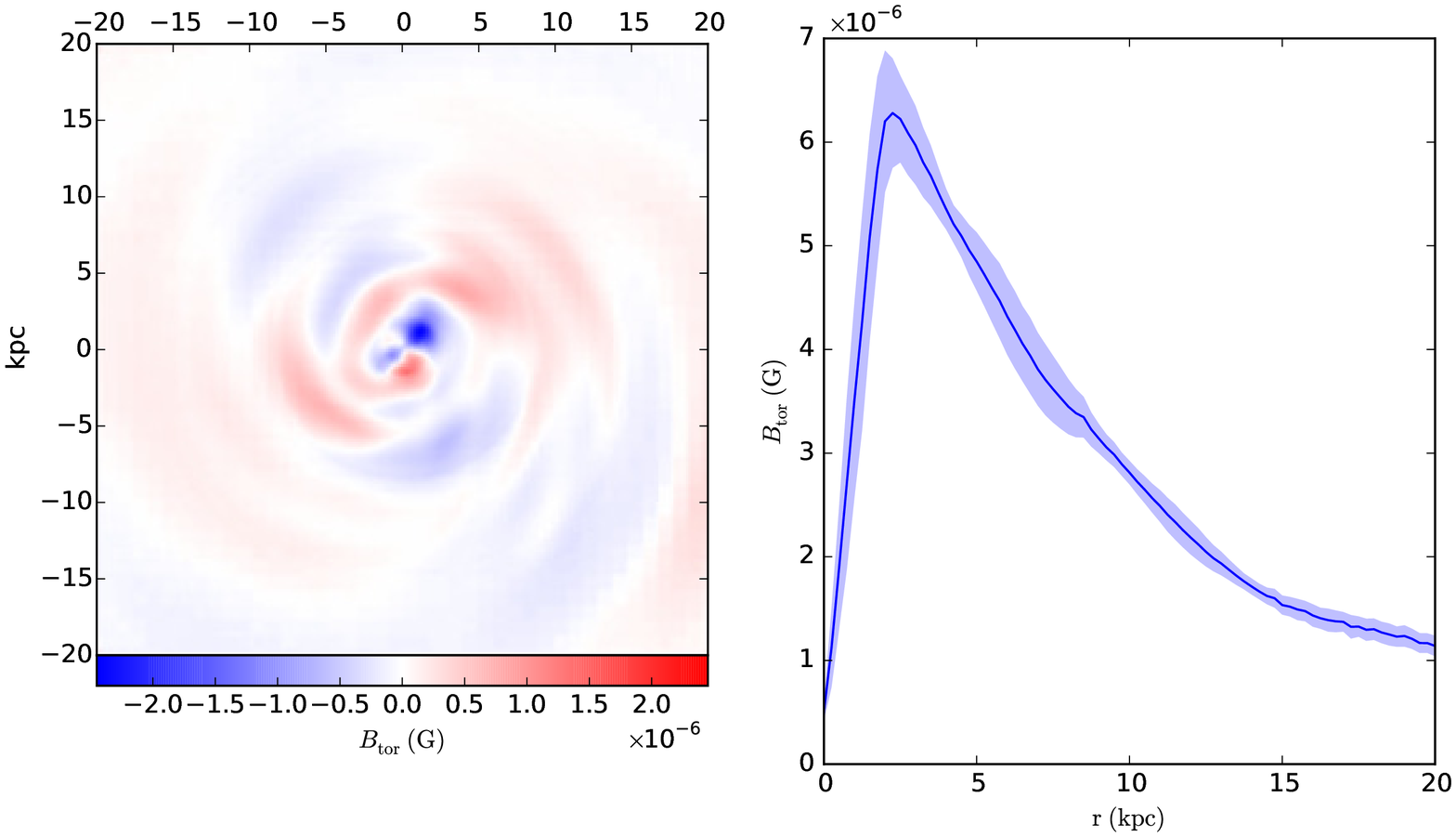}
\includegraphics[scale=0.65]{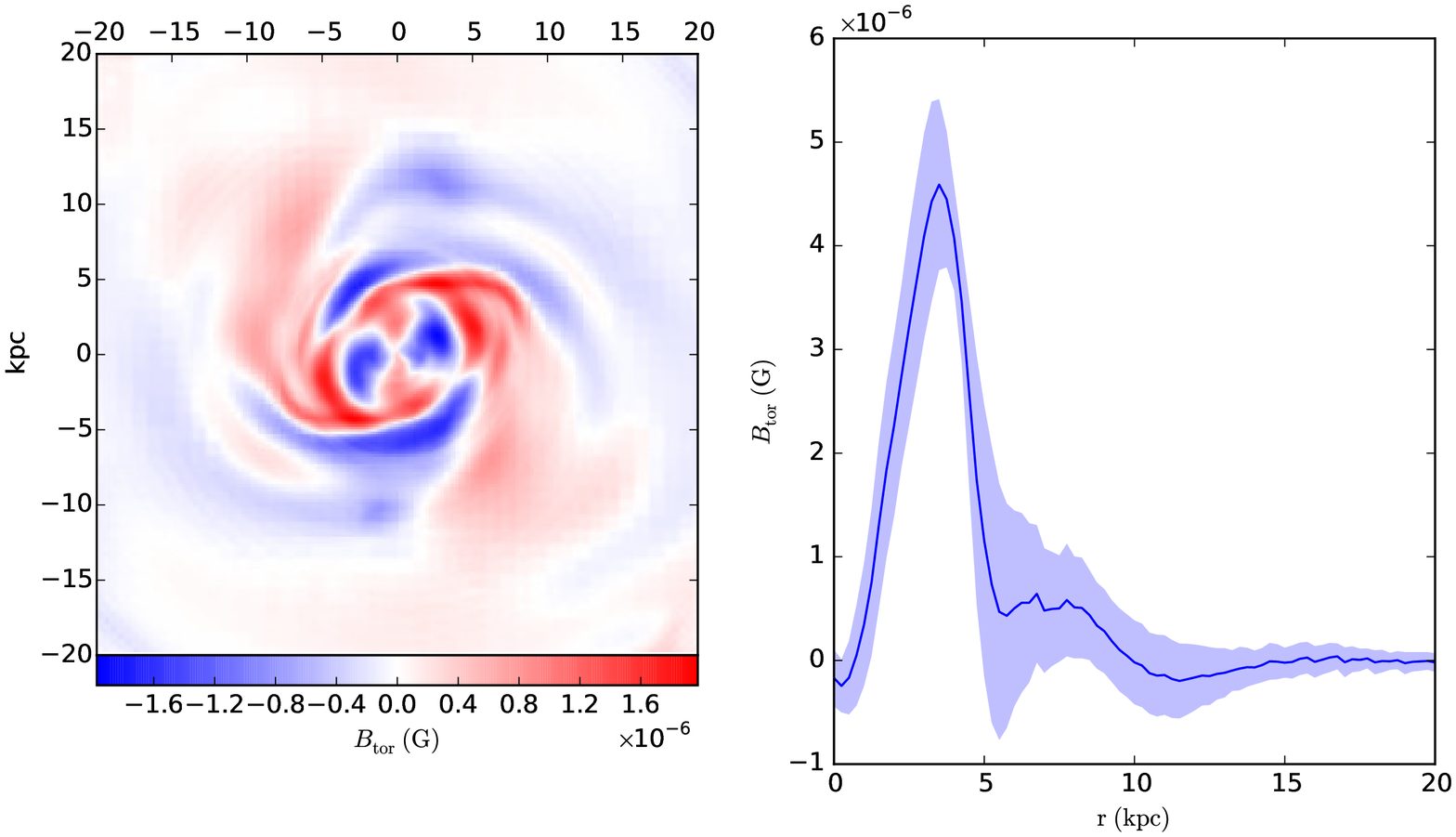}
\caption{{Magnetic field structure before the merger (at $3$~Gyrs) and after the merger (at $5$~Gyrs, , i.e. about $0.6$~Gyrs after the merger of the core).} Left: Projection of the fluctuation of the toroidal magnetic field strength (local toroidal field minus mean toroidal field at the same radius) in the central $40$~kpc. Right: Radial profile of the toroidal magnetic field as a function of radius, including its rms distribution. {Both plots refer to the reference run A.}}
\label{Btor}
\end{center}
\end{figure*}

\subsection{Reference run}
The global evolution in our simulation box is shown in Fig.~\ref{global} using density slices in the x-y plane. The isolated disks initially move in opposite directions along the x-axis, but start approaching each other along the y-direction due to their gravitational interaction. After $1$~Gyr, the low-density gas ($\sim10^{-27}$~g~cm$^{-3}$) at the very outskirts of the galaxies starts interacting, leading to the development of shocks. The interactions become more intense while the cores of the disks are approaching each other. After $4$~Gyrs, the interaction extends to higher-density gas components ($\sim10^{-25}$~g~cm$^{-3}$), leading to the formation of a disk around the galaxy cores, including signs of self-gravitational instabilities. The cores are clearly merged after $5$~Gyrs, and the disk surrounding the central core shows spiral arm features characteristic for self-gravitating disks.

The evolution closer to the merger event is shown in Fig.~\ref{sigma}, showing density projections at different times between $3.25$~Gyrs and $5.6$~Gyrs. At $3.25$~Gyrs, the cores of the disks are still separated, while the surrounding gas starts to merge and form a joint structure surrounding the cores. At $3.75$~Gyrs, the cores have reached a separation of less than $10$~kpc. When reaching this separation, the cores have however still a significant amount of kinetic energy, so that the separation for increases again toward $\sim4$~Gyrs. The overall process is however strongly damped by the hydrodynamical interaction, and the cores strongly approach each other {again} at $4.4$~Gyrs, leading to the final merger event. At $3.75-4$~Gyrs, the gas surrounding the cores shows bar-like features and signs of strong gravitational instability, which develop into spiral arm features after the merger. 

The corresponding evolution of the magnetic energy density is given in Fig.~\ref{Bdens}, which is closely related to the evolution of the density structure. For a comparison  with observational data, in particular the merger sequence compiled by \citet{Drzazga11}, we calculate the magnetic field strength averaged over different scales during the evolution of the merger. In particular, we determine the average field strength within $5$~kpc, within $25$~kpc and in the region between $5$ and $25$~kpc. These regions can be roughly associated with the magnetic field strengths provided by \citet{Drzazga11}, who were considering the average magnetic field strength in the galaxy center ($\sim6.5$~kpc), in the whole galaxy and in the outer parts of the galaxy. The regions employed for the averaging procedure are denoted with circles in Fig.~\ref{Bdens}, and the evolution of the resulting magnetic field strengths as a function of time is given in Fig.~\ref{av}. {The circles are centered onto the density peaks of the simulation. Especially in cases of two disks with equal mass, the position of the peak may then jump from one disk to another. As the situation is highly symmetric, the latter is however not relevant for the determination of the averaged magnetic field strength on the different scales, giving rise to a rather smooth evolution as shown in Fig.~\ref{av}. In case of disks with unequal mass, we have checked that the density peak remains on the disk with the higher mass.}

{For the magnetic field averages on all three scales explored here}, clear peaks can be recognized at about $3.75$~Gyrs and $\sim4.4$~Gyrs. While the peak at $4.4$~Gyrs corresponds to the merger of the galaxy cores, the cores are still well-separated at $\sim3.75$~Gyrs. The latter can be understood from a comparison with Fig.~\ref{Bdens}. At a time of $3.75$~Gyrs, the galaxy cores are still separated, but close enough that both galaxy cores fall into {the characteristic scale of $25$~kpc}. This naturally explains the peak in the field strength averaged over the entire galaxy, as an additional strong contribution from the second core is included in the average. Similarly, this core will also contribute to the magnetic field strength in the outer region{, corresponding to distances between $5$ and $25$~kpc from the center} of the first core. In addition, however, also the magnetic field strength within the central $5$~kpc is  enhanced by a similar factor. As the cores are still clearly separated, the latter is not due to a geometric effect. Rather, due to the tidal torques resulting from the presence of the additional core, the angular momentum transport within the galaxy is enhanced, increasing the density and the magnetic field strength within the central region,  explaining the peak in the central region of the disk. This is consistent also with the density evolution found in Fig.~\ref{sigma}.

For the peak at $4.4$~Gyrs, both cores are again within the radius of $25$~kpc, explaining the enhancement both on the scale of the galaxy and its outer regions. During that time, also the two cores actually merge, providing again a physical enhancement of the magnetic field strength on scales of $\sim5$~kpc. We note that the second peak is smaller than the first one. This is at least partly due to the fact that the gas in the galaxy, which has come very close during the merger event, starts encompassing again a somewhat larger surface area, as not all the kinetic energy has been dissipated by the shocks, and some of it leads now to the formation of a more well-defined disk due to the presence of angular momentum. The latter leads also to a redistribution of the magnetic energy.

While the magnetic field strength before the merger is well-aligned with the rotation of the galaxy and predominantly has a toroidal component, the latter changes significantly due to the merger event. In Fig.~\ref{Btor}, we show on the left-hand side the fluctuation of the toroidal magnetic field component, defined as its local value at a given position minus the mean toroidal field strength at the same radius, in a projection of the central $40$~kpc {both at $3$~Gyrs well before the merger, as well as at a time of} $5$~Gyrs, i.e. about $0.6$~ Gyrs after the merger of the central cores. {At $3$~Gyrs, the initial profile of the magnetic field strength (resulting from equipartition with a fixed fraction $\epsilon_{\rm mag}$ of the thermal energy) is still mostly intact, and fluctuations in the toroidal component are generally small. After the merger, the radial profile of the toroidal field is strongly disrupted and influenced by the presence of spiral arms, and the scatter in the toroidal component is significantly enhanced, with an rms deviation that is comparable to the mean.}

The latter is further examined in Fig.~\ref{power}, where we show the power spectrum of the  magnetic field in our simulation as a time sequence. One can clearly recognize two peaks both at $3.75$~Gyrs and $4.4$~Gyrs, reflecting the additional power on small scales during the merger. While these peaks disappear after the merger, one still notices that a considerable amount of power has shifted to a smaller scale. In addition, the overall power in the field strength is decreasing after the merger, reflecting the behavior found also for the average strength of the magnetic field. {The latter is due to a slight expansion of the disk that we observe in all simulations after the merger, thus causing a spread both of the mass as well as the magnetic energy over a larger volume. We expect that an external potential due to the presence of a stellar and dark matter component would help to stabilize the size of the gaseous disks. Even in this case, a relaxation of the disk after the merger would however be expected due to the release of gravitational energy.}

\subsection{Parameter variations}
\begin{figure*}[htbp]
\begin{center}
\includegraphics[scale=0.55]{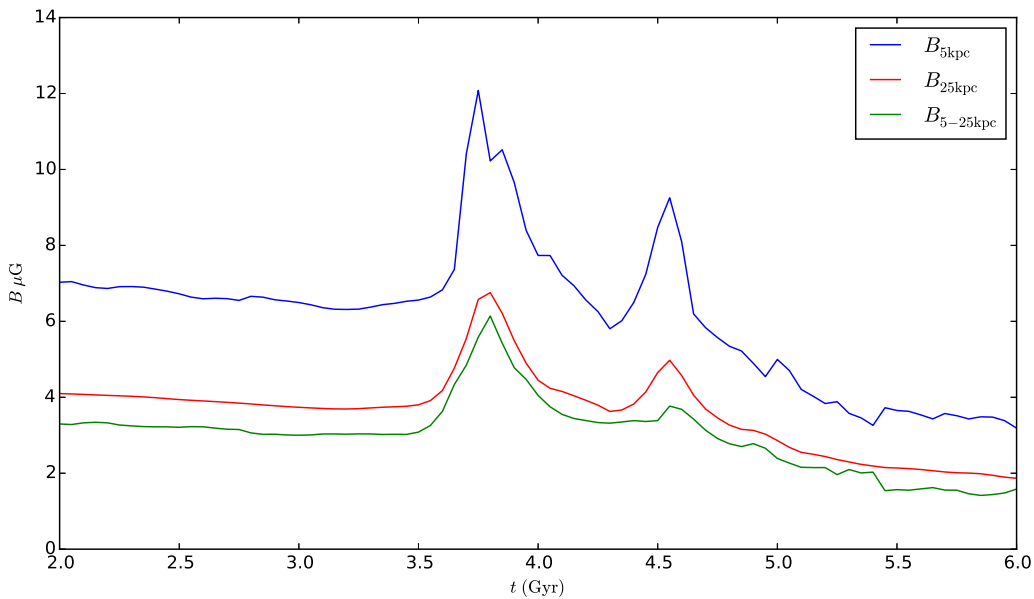}
\includegraphics[scale=0.38]{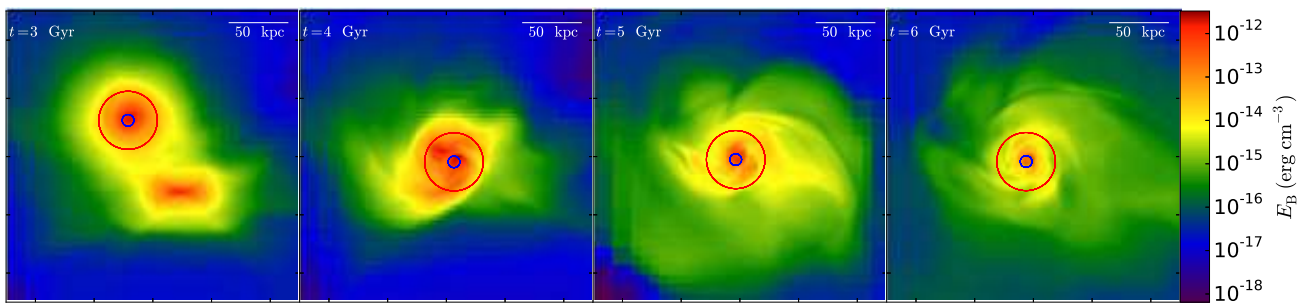}
\caption{Comparison run with a relative disk orientation of $90^\circ$ {(run B)}. Top: Time evolution of the mean magnetic field strength on different scales. Bottom: Projection of the magnetic pressure at different times.}
\label{perp}
\end{center}
\end{figure*}

\begin{figure*}[htbp]
\begin{center}
\includegraphics[scale=0.55]{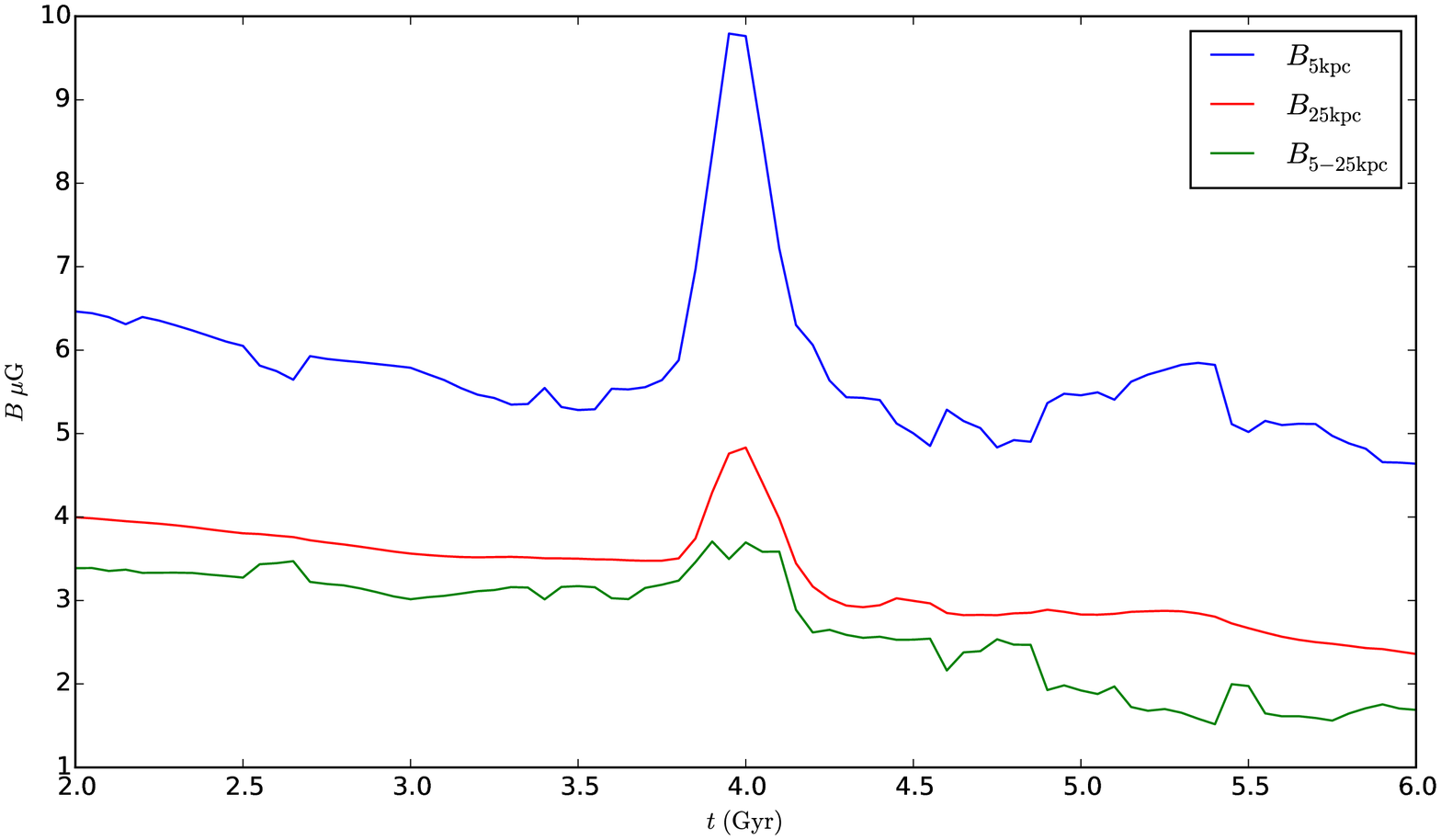}
\includegraphics[scale=0.38]{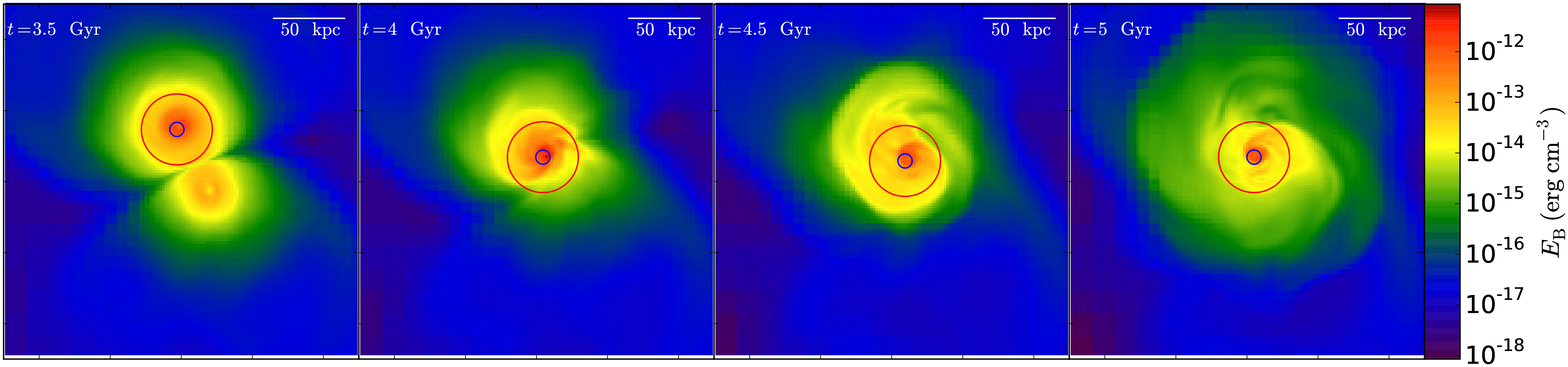}
\caption{Comparison run with a mass ratio of $1:3$ {(run C)}. Top: Time evolution of the mean magnetic field strength on different scales. Bottom: Projection of the magnetic pressure at different times.}
\label{ratio}
\end{center}
\end{figure*}

\begin{figure*}[htbp]
\begin{center}
\includegraphics[scale=0.5]{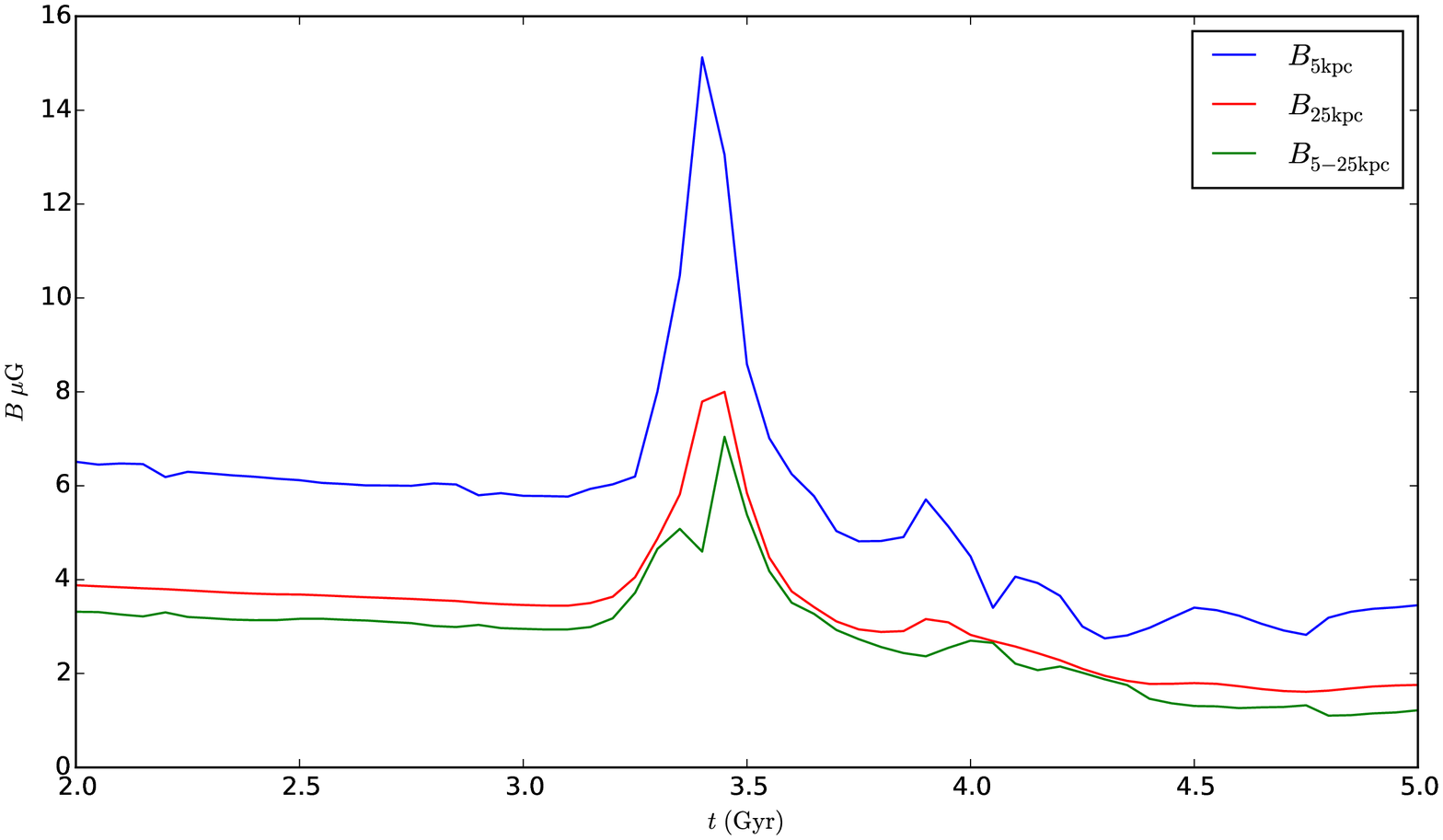}
\includegraphics[scale=0.5]{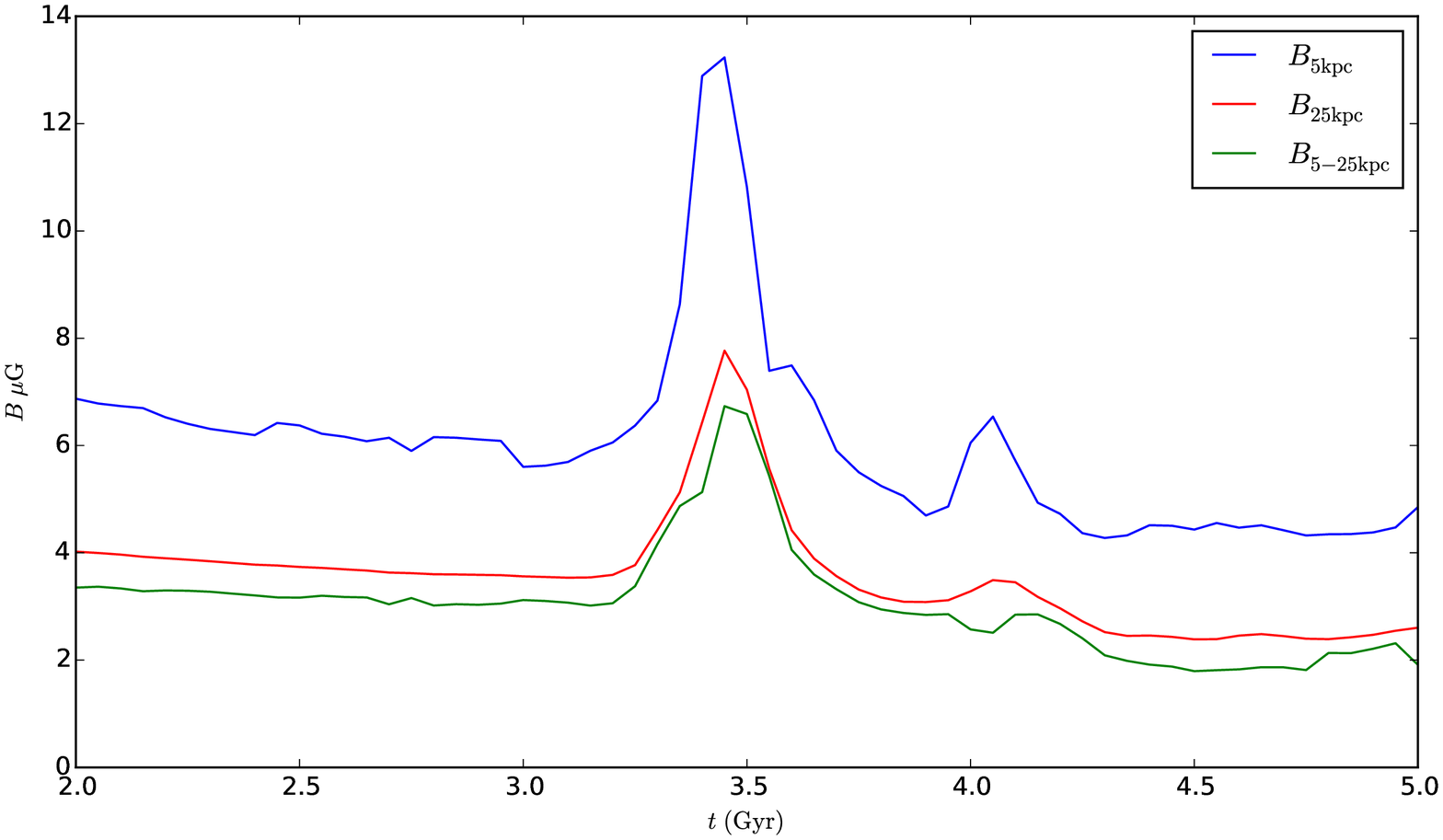}
\caption{Comparison runs with a lower relative velocity between the disks. We focus here on the time evolution of the magnetic field strength averaged over different length scales. Top: Relative velocity of $10$~km/s {(run D)}. Bottom: Relative velocity of $15$~km/s {(run E)}.}
\label{velocity}
\end{center}
\end{figure*}

{In order to at least qualitatively understand how the results shown here depend on the adopted configuration, we have varied the choice of the orbit, the relative disk orientation and the merger mass ratio. As a first interesting case, we present a simulation in Fig.~\ref{perp} with a relative disk orientation of $90^\circ$ {(run B)}. In this case, the global dynamics before the merger event change only to a minor degree. In particular, there are still two events where both disks fit into the beam, showing an apparent amplification in an average over galactic scales.

To explore the effect of varying the mass ratio, while keeping our focus on major mergers, i.e. neglecting minor events that can be considered as small perturbations of the main galaxy, we also show a simulation with mass ratio $3:1$ in Fig.~\ref{ratio} {(run C)}, where the mass of the more massive disk is unchanged. This changes the dynamics sufficiently to lead to a more rapid merger event. As a result, the disks merge directly after the first encounter, providing only one peak in the large-scale averaged magnetic field strength, and without orbiting the larger galaxy another time.

\begin{figure*}[htbp]
\begin{center}
\includegraphics[scale=0.7]{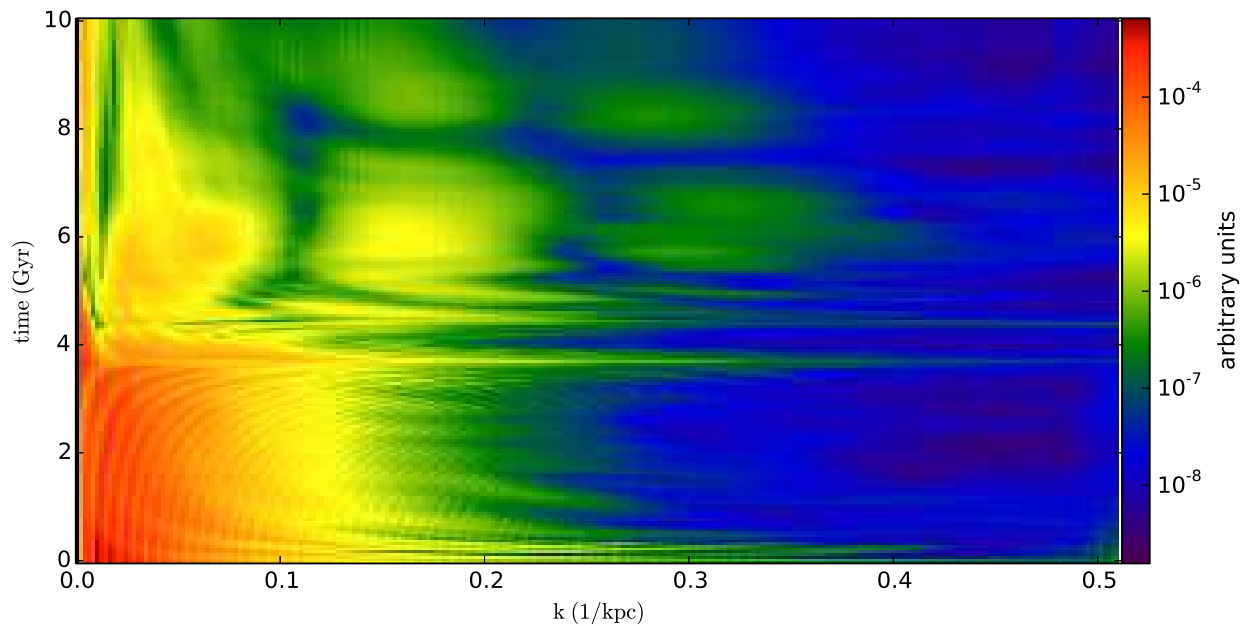}
\caption{Power spectrum of the magnetic field in the simulation box as a sequence of time {(run A)}. While the magnetic energy is initially localized on larger scales due to the coherent toroidal field, one can see a distinct peak at $3.6$~Gyrs and $\sim4.4$~Gyrs due to the generation of fluctuations on  smaller scales in particular due to the decreasing separation of the central cores. While these peaks disappear again after the merger, a noticeable amount of power is nevertheless shifted to smaller scales.}
\label{power}
\end{center}
\end{figure*}

To explore changes in the orbit, we also present two simulations where we have changed the relative initial velocity, as shown in Fig.~\ref{velocity}. These cases correspond to initial relative velocities of $10$~km/s {(run D)} and $15$~km/s {(run E)}, respectively. In both cases, the merger of the disks significantly accelerates, and we see essentially one peak in the time evolution of the average magnetic field strength. The bottom panel of the figure, corresponding to the relative velocity of $15$~km/s, still shows a minor hint of a potential second peak, as a result of an incomplete initial merging process. We however note that this feature is not comparable in size to what we see in our reference run. 

While of course the sample of configurations adopted here is still limited, the occurence of one or two peaks { in the magnetic field strength evolution} as a result of geometric projection events is a generic phenomenon we can infer from these simulations. The implications of these results should thus be considered in the interpretation of data from larger beams. {As a possible way to consider the geometric projection effects described here, we suggest to normalize the inferred strength of the magnetic field with respect to the size of the gaseous component, providing a better measure of the magnetic field strength in the galaxy rather independently from the size of the magnetized region.}
}

\section{Discussion and conclusions}\label{discussion}

To explore the evolution of magnetic fields during merger events, we present a toy model to simulate the evolution of magnetic fields during the mergers. For this purpose, we have modeled the interaction of two gaseous magnetized disks to explore how the interaction of the gas changes the structure of the magnetic field. In these idealized simulations, we have neglected the presence of a stellar component and a dark matter halo. {While the neglect of the dark matter halo can certainly have an impact on the orbit and change the dynamics, we have also seen in our sample of simulations that a set of characteristic features is rather insensitive to the details of the orbit. We describe in more detail below how part of the results may change after the inclusion of dark matter halos.}

In spite of the simplicity of our model, we are able to reproduce central characteristics of the magnetic field evolution in the merger sequence provided by \citet{Drzazga11}. In particular, they have shown a characteristic peak in the average field strength of the galaxy, its central regions and also in the outer parts of the galaxy when plotted against the interaction parameter. For comparison, we have followed the time evolution of the magnetic field in the central $5$~kpc of the galaxy, within a $25$~kpc radius and in the region between $5$~kpc and $25$~kpc. We expect the latter to approximately resemble the three cases distinguished by \citet{Drzazga11}. {In our reference run}, we find two characteristic peaks in these quantities during the merger event. The peak in the average field strength within the galaxy and its outer region can be naturally explained by geometrical effects when the core of the second disk falls into the region  covered by the $25$~kpc radius. Within the central $5$~kpc, both the magnetic field strength and the central density are actually enhanced, due to the increased angular momentum transport induced by the tidal torques within the configuration. The first peak occurs when the two cores for the first time come very close, but without merging, while the second peak occurs when the two cores are in fact about to merge.  {We however note that the occurence of two peaks is not critical and indeed one peak in the averaged field strength, as observed in other simulations, is indeed enough to explain the sequence derived by \citet{Drzazga11}.}

Despite their simplifications, our simulations can thus naturally explain the occurence of a peak in the magnetic field strength during the merger. We note that the time sequence produced in the simulation may not correspond one to one to the merger sequence produced by \citet{Drzazga11}, and in particular the interaction parameter may not uniquely correspond to a single time in the simulation, but rather to the characteristic times when the central cores come very close. Concerning the physical interpretation, our  results show that one must carefully distinguish between geometrical effects on the one hand, as the average field strength will increase when the core of one galaxy enters the outer parts of another one{, and physical changes in the magnetic field strength in particular in the center of the galaxy, either due to density enhancements or potentially due to dynamo effects.} High-resolution observations to explore the central regions of merging galaxies are therefore particularly valuable to further explore the increase of the magnetic field strength during the merger along with its physical origin.

We have checked that the results presented here do not strongly depend on the parameters in the simulations, such as the mass of the galaxy, their relative velocity or the initial magnetic field strength. In some cases, we found the emergence of only one peak rather than two, depending on the detailed evolution of the final merger. {Such behavior tends to occur when the relative velocities are lower between the galaxies, leading to a more rapid merger event. In this case, there will be only one event where both disks fall into the observed beam, leading to two contributions. This also confirms that geometrical effects play the dominant role in case of large beams, and only high-resolution observations can help to assess the physical behavior of the magnetic fields.}

{For realistic simulations, a dark matter halo should be included for both galaxies. To zero order, the latter will enhance both the gravitational and the inertial mass. While the baryons in the disks were strongly interacting leading to a relevant amount of friction, the latter will be reduced for the mass in dark matter. However, we note that even the dark matter halos will experience dynamical friction while moving through the circumgalactic material, including both baryonic and dark matter. While a precise orbit can only be obtained via dedicated simulations for each specific configuration, we expect that overall friction effects will be less relevant in a dark matter dominated regime, suggesting that the final merger may be somewhat delayed. The latter may affect the overall probability to find a specific galaxy in a certain merger stage, while the resulting geometrical projection effects are likely the same.}

As we noted in the introduction, previous investigations concerning the evolution of magnetic fields during galaxy mergers have been predominantly pursued using MHD-SPH simulations \citep[e.g.][]{Kotarba10, Kotarba11, Geng12}, or in the 2D-approximation using a grid-based approach \citep{Moss14}. Even the number of studies of isolated galaxies including magnetic fields is still limited. In this respect, we note the simulations pursued by \citet{Wang09} exploring galaxy formation in a cosmological context, where they neglected stellar feedback. More recently, \citet{Pakmor13} pursued simulations of isolated disk galaxies including magnetic fields and stellar feedback using a moving-mesh approach. Magnetic field amplification in isolated galaxies is further explored in the 2D approximation \citep{Moss15} or through the simulation of local volumes \citep{Gressel08}. The interaction of the dynamo process with cosmic rays has further been explored by \citet{Hanasz09}. Hydrodynamical simulations of isolated galaxies have been pursued with the Enzo code by \citet[e.g.][]{Tasker09}, and gravitational instabilities in isolated disks have been explored by \citet{Lichtenberg15}.

While we have shown here that even simplified models can already reproduce some of the main features in the observed merger sequence, future simulations should  aim for a more detailed and more realistic modeling, accounting for the effects of feedback, and employing cosmological initial conditions. In the context of isolated disk galaxies, the latter has been achieved by \citet{Wang09}, \citet{Latif14} and \citet{Pakmor14}, but it is still outstanding in the context of galaxy mergers. An additional problem is the limited resolution in numerical simulations, implying that galaxy-scale simulations will never be able to resolve turbulence and small-scale magnetic fields on all relevant scales. In the magneto-hydrodynamical equations, it may therefore be necessary to adopt nonlinear closures to explore the resulting dynamical implications \citep{Grete15}. The latter was shown to be relevant in hydrodynamical simulations of isolated galaxies \citep[e.g.][]{Braun14, Braun15}, and we expect similar effects to occur in MHD simulations.

\begin{acknowledgements}
We thank for fruitful discussions with Muhammad Latif concerning the MHD implementation in Enzo. The simulation results are analyzed using the visualization toolkit for astrophysical data YT \citep{Turk11}.  KR acknowledges financial support by the International Max Planck Research School for Solar System Science at the University of G\"ottingen. DRGS thanks for funding through Fondecyt regular (project code 1161247) and through the ''Concurso Proyectos Internacionales de Investigaci\'on, Convocatoria 2015'' (project code PII20150171).
\end{acknowledgements}

\end{document}